# Autonomous Cyber Defense in Connected Vehicles: A Multi-Agent Approach to V2X Security

Krishna Teja Medam
*Dept. of Electrical Engineering & Computer Engineering, and Computer Science*
University of Detroit Mercy
Detroit, MI, USA
kmedam@udmercy.edu

***Abstract*- A connected vehicle has roughly 100 milliseconds to decide whether an incoming Basic Safety Message is real or fabricated. If a false emergency braking alert reaches the planning pipeline in time, the car brakes. That is a safety failure triggered by a security failure, and existing intrusion detection systems are not designed to handle that coupling. They operate per vehicle, per message, with static rules - blind to attack patterns that only emerge across a fleet or over time, and blind to the fundamental tension between dropping a suspicious message and dropping a real emergency alert. We propose a three-tier multi-agent architecture that treats this timing constraint as a hard design requirement, not a performance target. At the vehicle level, an onboard agent classifies each incoming V2X message into one of four actions - Accept, Drop, Quarantine, or Escalate - within a 10-millisecond budget. The classifier is deliberately biased toward Escalate when uncertain, passing ambiguous cases to the roadside edge agent rather than risking a dropped legitimate alert. The edge agent operates across a roadside unit zone with a 50-millisecond budget, fusing threat assessments from multiple vehicles and resolving safety-security conflicts using complementary sensor observations. The cloud tier refines detection models through Byzantine fault-tolerant federated learning and redistributes updated weights to the fleet. Every timing constraint derives directly from the 100-millisecond Basic Safety Message cycles mandated by SAE J2735 and ETSI EN 302 637-2. This paper does not present simulation results or empirical validation - the contribution is architectural. We argue that no existing framework simultaneously assigns standards-grounded latency budgets to all three deployment tiers while treating safety-security conflict resolution as a first-class design constraint. Remaining open problems - adversarial model poisoning at the edge and the absence of regulatory frameworks for autonomous security response - are discussed as concrete directions for future work.**

## I. INTRODUCTION

Modern transportation infrastructure is quietly undergoing one of the most consequential security transitions in its history. Vehicles once isolated by physics - separated by metal, glass, and distance - are now continuously exchanging messages with each other, with roadside infrastructure, and with cloud services. The Vehicle-to-Everything (V2X) communication paradigm makes this possible, enabling connected vehicles to share Basic Safety Messages (BSMs), signal phase and timing data, road hazard warnings, and intersection geometry. SAE J2735 defines the message set dictionary governing this exchange, while its companion standard J2945/1 mandates a 100-millisecond transmission cycle - ten messages per second - for safety-critical BSMs [17]. The safety benefits are real: cooperative awareness reduces rear-end collisions, enables platooning, and supports automated emergency braking at response times no human driver can match.

The security implications are equally real, and considerably less discussed.

Every message a vehicle trusts is a message that could have been fabricated. A BSM claiming a vehicle is braking hard ahead will cause an automated driving system to brake, regardless of whether that vehicle exists. A spoofed GPS coordinate embedded in a Map Data (MAP) message can redirect routing decisions. Sybil attacks - in which a single adversary generates multiple false vehicle identities - are a demonstrated threat in deployed V2X environments, with experimental evidence confirming their feasibility against real vehicular communication channels [5]. The 100-millisecond BSM cycle that makes cooperative driving possible is the same window within which a fabricated message can reach an automated braking system and trigger a physical response [17].

The security mechanisms currently deployed in V2X networks are not designed for this threat model. The Security Credential Management System (SCMS) and IEEE 1609.2 provide certificate-based message authentication that verifies a message came from a legitimate enrolled device [10]. What they cannot verify is whether the message content is truthful. A compromised vehicle with valid credentials can inject semantically plausible false data indefinitely. Intrusion detection systems represent a complementary layer, but existing approaches are detection-only: they flag anomalies for offline review without triggering any autonomous protective action [7], and they operate within a single vehicle, unable to observe the fleet-wide patterns that distinguish a coordinated multi-vehicle attack from a localized sensor fault [13].

This mismatch between threat and defense has three structural causes. First, existing systems produce alerts but delegate response to human operators or static rule engines - a detection-without-response gap that is incompatible with the 100-millisecond V2X timing constraint within which a threat must be neutralized [16]. Second, per-vehicle IDS architectures are architecturally blind to coordinated attacks: the evidence that a Sybil campaign is underway exists across vehicles simultaneously, not within any one of them [13]. Third, the safety-security tradeoff - the risk that aggressively filtering suspicious messages suppresses a legitimate emergency alert - has received almost no attention as a formal design constraint in existing V2X security frameworks [4]. A system that drops a real emergency braking message to prevent a spoofed one from getting through has not improved safety; it has traded one failure mode for another.

This paper proposes a three-tier multi-agent framework that addresses these limitations by treating the V2X timing budget as an architectural constraint rather than a performance aspiration. An onboard in-vehicle agent performs four-class threat classification - Accept, Drop, Quarantine, or Escalate - within a 10-millisecond budget grounded in the SAE J2735 100-millisecond BSM cycle [17]. A roadside unit (RSU) edge agent operates across a geographic zone with a 50-millisecond budget, correlating threat signals from multiple vehicles and resolving safety-security conflicts using cross-modal sensor observations - a technique grounded in spatiotemporal consistency research [11]. A cloud agent manages fleet-wide model refinement through Byzantine fault-tolerant federated learning, drawing on decentralized FL methods developed for autonomous vehicle environments [12], and distributes updated detection weights to the fleet without centralizing raw vehicle data.

The contribution of this paper is not the invention of any individual component. Cross-modal sensor fusion for spoofing detection [11], Byzantine-robust federated learning for autonomous vehicles [12], and distributed cooperative misbehavior detection [13] each have prior treatments in the literature. The contribution is their unification into a coordinated architecture with explicit agent role definitions, standards-derived latency constraints, and safety-security conflict resolution as a first-class design requirement. The closest prior work integrates federated learning, graph neural networks, and reinforcement learning for V2X defense but reports an 18-millisecond response latency, defines no tier-based agent roles, and does not address safety-security conflict resolution [4] - falling short of the sub-10-millisecond Tier 1 budget derived from SAE J2735 BSM transmission cycles [17]. To our knowledge, no existing framework simultaneously satisfies all three constraints.

The remainder of this paper is organized as follows. Section II establishes the V2X communication background and formalizes the threat model. Section III reviews the relevant literature and identifies seven structural gaps the proposed framework addresses. Section IV presents the three-tier architecture in detail. Section V analyzes the framework's security coverage against the threat model. Section VI discusses open challenges. Section VII concludes.

# II. Background and Threat Model

## A. V2X Communication Architecture

Vehicle-to-Everything communication covers four interaction modes: Vehicle-to-Vehicle (V2V), Vehicle-to-Infrastructure (V2I), Vehicle-to-Pedestrian (V2P), and Vehicle-to-Network (V2N). Two physical-layer technologies compete in today's V2X deployments. Dedicated Short-Range Communications (DSRC) runs on the 5.9 GHz band using IEEE 802.11p, offering low-latency peer-to-peer communication without any cellular dependency. Cellular V2X (C-V2X) uses the 3GPP PC5 sidelink for direct vehicle communication and the Uu interface for network-assisted communication over 4G or 5G. Both technologies carry the same standardized message types that safety applications depend on, and both share the same attack surface.

The message set for V2X safety communication comes from SAE J2735 [17]. Four message types matter most for this paper's threat model. Basic Safety Messages (BSMs) are broadcast by every vehicle at 10 Hz, carrying position, speed, heading, and brake status - the primary input to cooperative collision avoidance. Signal Phase and Timing (SPAT) messages come from roadside units at intersections, telling vehicles the state and timing of traffic signals. Map Data (MAP) messages describe intersection geometry and lane connectivity. Traveler Information Messages (TIMs) carry road condition and hazard alerts. In Europe, the Cooperative Awareness Message (CAM) serves the same role as the BSM, with ETSI EN 302 637-2 requiring a minimum inter-message interval of 100 milliseconds [18]. The Decentralized Environmental Notification Message (DENM) handles event-driven emergency notifications with zero delay tolerance [19].

## B. PKI and the Security Credential Management System

North America's primary V2X security mechanism is the Security Credential Management System (SCMS), built on the public key infrastructure defined by IEEE 1609.2 [10]. Every enrolled vehicle gets a set of pseudonymous certificates - short-lived credentials that authenticate message origin without exposing a persistent identity. Each BSM is digitally signed with the vehicle's current pseudonym certificate, and receivers verify the signature before acting on the message.

This architecture blocks certain attack classes outright. A vehicle outside SCMS enrollment cannot inject messages that pass signature verification. Replay attacks using captured messages can be caught through certificate expiry and timestamp validation. The pseudonym rotation mechanism - examined in the comparative analysis of North American and European PKI architectures in [9] - limits long-term vehicle tracking across locations.

What PKI cannot do is semantic verification. It has no way to tell whether the content of a signed message is true. A vehicle with valid credentials that has been compromised, or whose sensors have been tampered with, can broadcast false position, speed, or hazard data that passes every cryptographic check. That content-authenticity gap is the fundamental reason this paper proposes a misbehavior detection layer.

## C. Latency Budget Derivation

Every tier latency budget in this framework traces back to V2X standards rather than arbitrary design decisions - an important distinction for both technical credibility and regulatory alignment.

SAE J2735 defines the BSM message format and content [17]. Its companion standard, SAE J2945/1, mandates that BSMs be transmitted at 10 Hz, producing a 100-millisecond inter-message interval that anchors all tier latency budgets. ETSI EN 302 637-2 independently requires the same 100-millisecond minimum interval for CAMs [18]. The convergence of both global standards on the same cycle makes 100 milliseconds the universal timing window within which all V2X security processing must complete.

The Tier 1 in-vehicle agent budget of sub-10 milliseconds is one-tenth of that cycle. This leaves 90 milliseconds for downstream ADAS processing after security classification - a conservative margin that keeps security inspection from becoming a bottleneck in the automated driving pipeline. For event-driven DENMs, which carry the highest safety urgency and tolerate no transmission delay [19], this budget matters even more.

The Tier 2 RSU edge agent budget of 10–50 milliseconds comes from realistic 5G V2X deployment latencies reported in the literature [6]. Together, the Tier 1 and Tier 2 pipeline (sub-10ms plus 10–50ms) gives a worst-case total under 60 milliseconds, well within one BSM cycle. Tier 3 cloud processing operates outside real-time constraints entirely; federated model updates are distributed on a minutes-to-hours basis and do not touch the per-message decision pipeline.

### D. Threat Model

This paper covers four attack classes the proposed framework must detect and respond to autonomously.

*Sybil Attacks:* A Sybil attacker creates multiple fake vehicle identities, either by cloning legitimate credentials or exploiting gaps in certificate revocation propagation. By injecting BSMs from dozens or hundreds of phantom vehicles, the attacker can fabricate traffic jams, create false emergencies that trigger widespread braking, or dilute cooperative perception maps to hide a real hazard. Detection approaches that rely solely on PKI authentication have been shown to be insufficient against Sybil threats; physical-layer fingerprinting beyond certificate verification is required [5].

*GPS Spoofing*: A GPS spoofing attacker sends counterfeit satellite signals that make a target vehicle's receiver compute a false position. That false position flows into BSMs and MAP messages, causing the victim to report an incorrect location and potentially follow wrong routing or collision avoidance decisions. Detecting spoofing requires cross-checking reported position against independent physical sensor observations - the core idea behind the PhyScout spatiotemporal consistency approach [11], which motivates the Tier 1 cross-modal reasoning component of this framework.

*Replay Attacks:* A replay attacker captures legitimate V2X messages and retransmits them at a different time or location. A replayed emergency braking BSM can trigger unnecessary braking; a replayed SPAT message can create false signal state information. IEEE 1609.2 certificate timestamps and sequence numbers provide the primary defense, but replay detection must complete within the Tier 1 latency budget before the replayed message reaches the ADAS pipeline [10].

*Adversarial Machine Learning Poisoning:* As V2X security systems adopt learned detection models, a new attack surface appears: model poisoning. A coalition of compromised vehicles can feed corrupted model updates during federated learning aggregation, degrading detection accuracy across the fleet. Byzantine fault-tolerant aggregation methods - specifically Krum and Median aggregation - are designed to isolate and exclude poisoned updates before they affect the global model [12]. This threat class is unique in operating at the Tier 3 training pipeline rather than the real-time message processing pipeline.

### E. Attacker model

The framework assumes three attacker profiles. The insider attacker operates a vehicle enrolled in SCMS with valid credentials, letting it inject cryptographically signed false messages that pass PKI verification. The external radio adversary operates outside SCMS but within radio range, capable of GPS spoofing, signal jamming, and replay injection. The federated poisoning attacker controls one or more vehicles in the Tier 3 federated learning process, contributing corrupted model updates to degrade fleet-wide detection accuracy.

The framework does not address physical compromise of vehicle hardware, supply chain attacks on ECU firmware, or attacks that require prolonged physical access. These are out of scope for a V2X communication security framework and are addressed by orthogonal automotive cybersecurity standards, as examined in the gap analysis of ISO/SAE 21434 [14].

## III. Literature Review

The work relevant to autonomous V2X cyber defense spans four research communities that have largely developed in parallel: intrusion detection for vehicular networks, federated learning for distributed anomaly detection, PKI and misbehavior detection standards, and multi-agent AI for cybersecurity. Each community has made real progress within its own scope. The gaps in the literature are not gaps of individual capability - they are gaps of integration. This section reviews the relevant work through seven structural limitations that together motivate the proposed framework.

### A. Gap 1 - Detection Without Response

The dominant paradigm in vehicular intrusion detection is detection-only: systems flag anomalous behavior and generate alerts, then delegate all response decisions to human operators or external systems. Althunayyan et al. propose a two-stage hierarchical federated learning IDS using ANN and LSTM autoencoder components that achieves strong anomaly detection on CAN bus traffic [7]. Asaouer and Boubiche conduct a systematic survey of generative AI approaches - GANs, Transformers, VAEs, and diffusion models - for intra-vehicle IDS, showing that zero-day adaptability is achievable but confirming that every surveyed system stops at detection [15]. Kurunathan et al. come the closest to breaking this pattern, proposing an A2C reinforcement learning and LSTM combination for intrusion mitigation in software-defined vehicle V2X communications [16]. Even so, their system operates as a single agent without tier coordination, and its response logic does not account for the safety-security tradeoff introduced in Section II.

The detection-without-response gap has a specific consequence in V2X environments that does not apply to conventional network security: a detected threat that is not acted upon within the BSM cycle window is effectively unaddressed, because the message it failed to intercept has already reached the ADAS pipeline. Alerting a human operator is not a viable response at 100-millisecond timescales.

### B. Gap 2 - Single-Tier Architecture

No existing framework simultaneously spans the in-vehicle, RSU edge, and cloud deployment tiers with distinct agent roles at each level. Agrawal's federated graph learning framework for V2X defense [4] operates as a single logical system despite using distributed computation, with no explicit differentiation between per-vehicle processing, zone-wide aggregation, and fleet-wide model management. Kurunathan et al. [16] similarly treat the vehicle as a self-contained detection and response unit. The consequence is that design decisions made for one deployment context - typically the vehicle - are applied uniformly across contexts where latency constraints, data availability, and computational resources are fundamentally different. A sub-10ms in-vehicle decision and a fleet-wide federated learning update are not the same problem and should not share the same architectural layer.

### C. Gap 3 - Single-Vehicle Scope

Cooperative detection - aggregating observations across multiple vehicles to identify patterns invisible to any individual vehicle - is architecturally absent from most IDS designs. Althunayyan et al.'s hierarchical FL approach [7] does aggregate model updates across vehicles, but detection inference runs per-vehicle on local CAN bus data. Ben Mokhtar et al. address this gap most directly, proposing a distributed misbehavior detection system for cooperative driving networks that lets vehicles perform consensus-based detection without centralized infrastructure [13]. However, their system operates at the V2V layer only, with no integration into a broader multi-tier security architecture. Coordinated Sybil campaigns - where attack evidence is inherently distributed across observations from multiple vehicles - cannot be reliably detected by any per-vehicle system operating alone.

### D. Gap 4 - Static vs. Adaptive Defense

Rule-based and signature-based detection systems cannot generalize to attack variants they have not been trained on. Asaouer and Boubiche identify this as the central motivation for generative AI approaches in vehicular IDS [15], noting that traditional ML models degrade against novel attack patterns. Kurunathan et al. address adaptability through reinforcement learning, which learns response policies through interaction rather than fixed rules [16]. Federated learning offers a complementary path: by continuously refining detection models across the fleet, the system can incorporate observations of new attack patterns from any vehicle and distribute updated defenses to all vehicles. Chen et al.'s BDFL framework demonstrates the feasibility of this approach for autonomous vehicle environments [12], though without integration into a real-time per-vehicle decision pipeline.

### E. Gap 5 - Safety-Security Tradeoff Ignored

The safety-security tradeoff is the most consistently overlooked design constraint in the V2X security literature. It arises because V2X safety applications depend on receiving and acting on messages from surrounding vehicles and infrastructure. A security system that aggressively filters suspicious messages reduces attack surface but simultaneously increases the risk of suppressing legitimate safety-critical alerts. Dropping a genuine emergency braking BSM to prevent a spoofed one is not a security success - it is a safety failure of a different kind. Agrawal's framework [4] is explicitly noted as lacking safety-security tradeoff handling. No paper in the reviewed literature treats this tension as a formal design constraint with an explicit resolution mechanism. The proposed framework addresses this gap by assigning safety-security conflict resolution as the primary responsibility of the Tier 2 RSU edge agent, where cross-vehicle context is available to disambiguate ambiguous messages before a response decision is committed.

### F. Gap 6 - Absence of Cross-Modal Reasoning

Existing V2X security systems evaluate messages in isolation from the physical sensor data that could corroborate or contradict their content. A BSM reporting a vehicle at a specific position can be cross-checked against LIDAR returns, radar detections, and camera observations from the receiving vehicle - but no existing V2X IDS performs this check. Xu et al.'s PhyScout system [11] establishes the technical foundation for this capability, demonstrating that GPS spoofing attacks can be reliably detected by checking spatiotemporal consistency across heterogeneous physical sensors including GPS, LIDAR, camera, and IMU. PhyScout was developed for intra-vehicle sensor fusion, not V2X message validation, but the underlying principle - that physical reality constrains what messages can plausibly claim - transfers directly to the V2X context. The proposed framework instantiates this principle in the Tier 1 in-vehicle agent.

### G. Gap 7 - Explainability

As autonomous security agents make consequential real-time decisions - dropping messages, quarantining sources, escalating alerts - the interpretability of those decisions becomes a regulatory and operational requirement, not merely a research interest. Abdel Hakeem et al. address this gap in the federated IDS context, introducing SHAP-based explainability to FL-driven intrusion detection for connected vehicles [8]. Their work focuses on explaining detection outputs, not response decisions. Grimm et al.'s gap analysis of ISO/SAE 21434 [14] identifies incident handling and decision traceability as unresolved challenges in the automotive cybersecurity engineering lifecycle. No existing framework provides explainability at the agent decision level - the layer at which Accept, Drop, Quarantine, and Escalate outputs are generated - leaving a gap that must be addressed before autonomous V2X defense systems can achieve regulatory acceptance.

### H. Gap Synthesis – The Integration Novelty

Each gap identified above has received partial attention in the literature. Detection-only systems are well-developed [7], [15]. Autonomous response has been attempted at the single-agent level [16]. Cooperative detection exists at the V2V layer [13]. Adaptive learning through federated methods is established for autonomous vehicle environments [12]. Cross-modal sensor fusion for spoofing detection has been demonstrated [11]. Explainability has been introduced to FL-based IDS [8]. What does not exist is a framework that addresses all seven gaps simultaneously through a coordinated architecture.

The multi-agent AI literature provides the conceptual scaffolding. Raghavan and Schneier's analysis of agentic AI decision loops [1] establishes the OODA-based foundation for the in-vehicle agent's Observe-Orient-Decide-Act cycle.

Vinay et al.'s taxonomy of agentic AI in cybersecurity [2] positions coordinated multi-agent systems with defined roles as the current frontier - generation four and five in their classification - distinguishing them from earlier single-agent and pipeline-based approaches. The ACM framework for multi-agent threat mitigation in high-stakes systems [3] supports the architectural principle of assigning distinct responsibilities to agents operating at different abstraction levels.

The proposed framework sits at this intersection. It does not claim that any individual component is new. Its contribution is the unification of these capabilities into a coordinated architecture - with explicit role definitions, standards-derived latency budgets, and safety-security conflict resolution as a first-class constraint - into a single architecture that no prior work provides.

## IV. Proposed Three-Tier Multi-Agent Framework

### A. Architecture Overview

The proposed framework organizes autonomous V2X cyber defense across three deployment tiers, each with an explicit agent role, a standards-derived latency budget, and a defined scope of decision authority. This is not three independent systems running in parallel. It is a coordinated pipeline: Tier 1 handles every message in real time, Tier 2 resolves cases Tier 1 cannot confidently classify alone, and Tier 3 continuously improves the detection models that Tier 1 and Tier 2 depend on. Figure 1 depicts the proposed three-tier architecture.

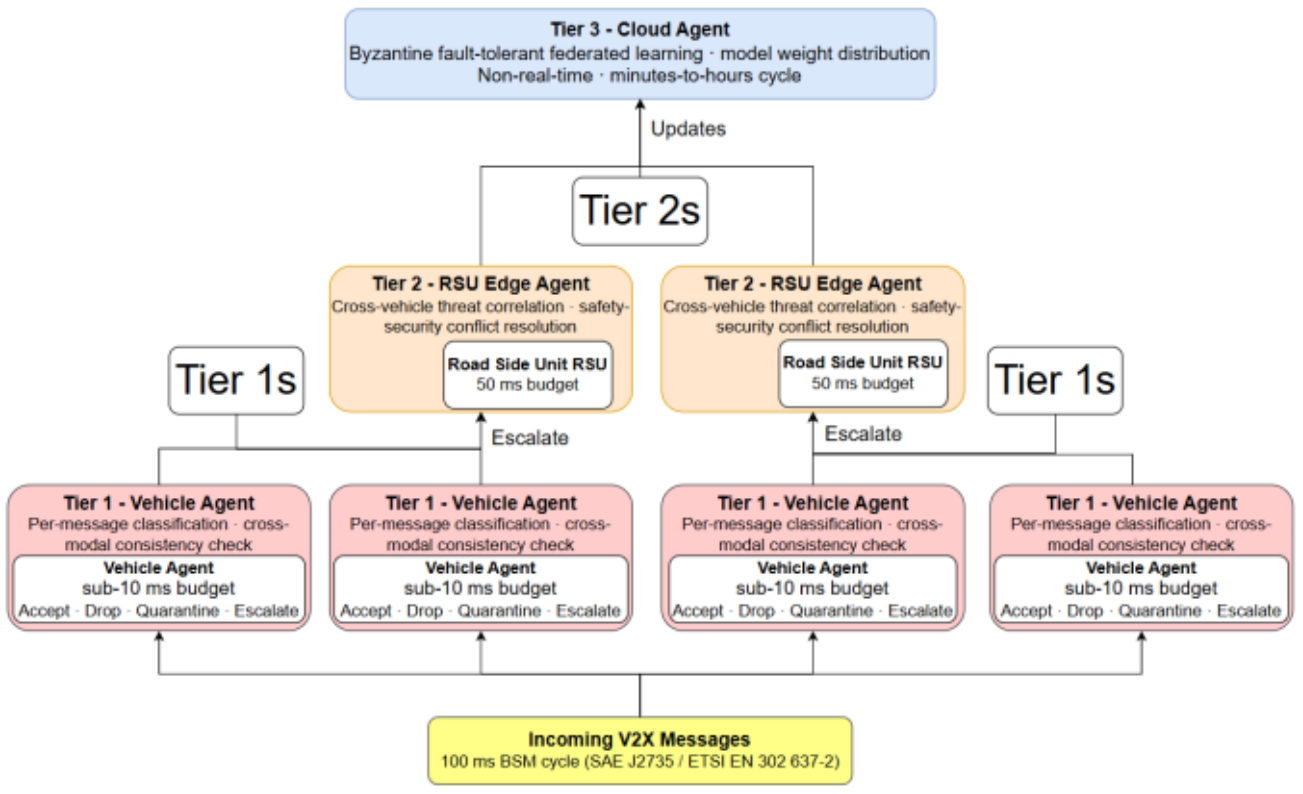

**Figure 1 Three-tier multi-agent architecture for autonomous V2X cyber defense**

The tier structure is architecturally forced by the V2X operating environment rather than chosen arbitrarily. A single in-vehicle agent cannot see the fleet-wide patterns needed to detect coordinated attacks - that requires aggregating observations across vehicles, which is a Tier 2 responsibility. A single RSU cannot manage fleet-wide model updates without becoming a centralized point of failure and a privacy liability - that requires a cloud layer with federated aggregation. And a cloud agent cannot meet the sub-10ms per-message classification requirement - that can only be satisfied by an onboard compute unit co-located with the V2X radio stack. Each tier exists because no other tier can perform its function within the constraints the V2X environment imposes.

The framework draws on four distinct bodies of prior work: the OODA loop model for agent decision architecture [1], spatiotemporal cross-modal consistency for in-vehicle spoofing detection [11], Byzantine fault-tolerant federated learning for fleet-wide model management [12] and distributed cooperative misbehavior detection for zone-wide threat correlation [13]. The contribution is their integration into a unified architecture with explicit inter-tier communication protocols, role boundaries, and a safety-security conflict resolution mechanism that no prior work provides.

### B. Tier 1 - In-Vehicle Agent

*Role and Scope:* The Tier 1 agent runs on the vehicle's onboard compute unit and is responsible for classifying every incoming V2X message before it reaches the ADAS pipeline. It is the only tier operating within the 100-millisecond BSM cycle constraint, and its sub-10ms budget leaves 90 milliseconds for downstream processing. It has no visibility into other vehicles' observations and makes no attempt to correlate across the fleet - those responsibilities belong to Tier 2.

*Decision Architecture:* The agent follows an OODA loop structure [1]: it Observes the incoming message and available sensor data, Orients by comparing message content against its current detection model and physical sensor readings, Decides on one of four classification outputs, and Acts by routing the message accordingly. The four outputs are:

- **Accept** - message passes all checks; forwarded to the ADAS pipeline immediately.
- **Drop** - message fails hard criteria (invalid certificate, expired timestamp, sequence number violation per IEEE 1609.2 [10]); discarded without forwarding.
- **Quarantine** - message is anomalous but not definitively malicious; held pending Tier 2 review; a safe default assumption is applied to the ADAS pipeline in the interim.
- **Escalate** - message cannot be classified confidently within the latency budget; forwarded to Tier 2 with full context for zone-wide assessment.

The classifier is deliberately biased toward Escalate over Drop when uncertainty is high. Dropping an unclassified message, risks suppressing a legitimate safety alert; escalating passes the decision to a tier with more context. This bias is the operational implementation of the safety-security tradeoff principle established in Section II. The complete Tier 1 decision pipeline, including early-exit escalation paths, is illustrated in Figure 2.

*Cross-Modal Consistency Check*: Before classifying a BSM reporting a specific vehicle position, the Tier 1 agent cross-checks the reported position against available onboard sensor observations - LIDAR point clouds, radar returns, and camera detections within the claimed direction and distance. This approach is grounded in the spatiotemporal consistency principle demonstrated by PhyScout [11]: physical sensor data and V2X message content must be consistent with each other within the constraints of physical reality. A BSM claiming a vehicle is braking at a position where no radar or LIDAR return exists within expected range triggers an Escalate output rather than an Accept. The agent does not need certainty - it needs enough inconsistency to justify escalation.

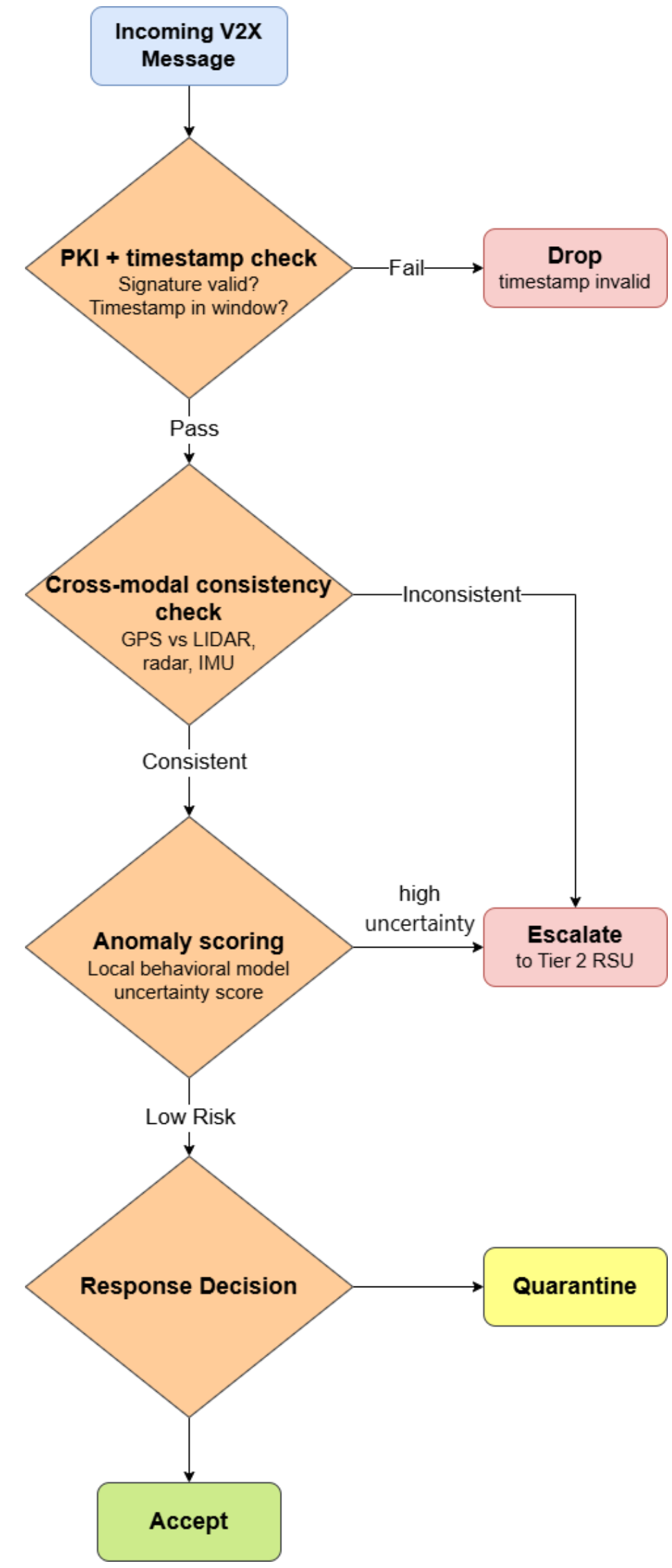

**Figure 2 Per-message decision pipeline at Tier 1 with escalation bias for ambiguous V2X messages.**

*Anomaly Scoring*: After the cross-modal check, the message proceeds to anomaly scoring against a local behavioral model, which encodes the vehicle's learned expectation of normal V2X traffic patterns: message frequency, plausible vehicle densities for the current road type, and typical kinematic profiles for the reported vehicle class. Each incoming message is compared against the local baseline of traffic patterns to produce a per-message uncertainty score that quantifies how far the message deviates from expected behavior. A message whose attributes are consistent with the local model receives a low uncertainty score and proceeds to the response decision stage. A message whose attributes are anomalous, but whose physical consistency could not be definitively resolved by the cross-modal check, receives a high uncertainty score that triggers an Escalate output. This step acknowledges a structural limitation: the local model captures one vehicle's observations, not the fleet's. A Sybil campaign injecting individually plausible BSMs will score low uncertainty at Tier 1 regardless of aggregate maliciousness [13]. The architectural response is deliberate escalation bias rather than a more complex local model: when local context is insufficient, the anomaly score and consistency evidence are passed to Tier 2 where cross-vehicle correlation is available [13], [4]. The Tier 1 budget constrains scoring to lightweight statistical comparison against recent observation history; deeper pattern analysis is deferred to higher tiers by design.

*Graceful Degradation*: When no RSU is available - in rural corridors, during maintenance, or early in deployment - the Tier 1 agent operates in standalone mode. Escalated messages that cannot reach Tier 2 are handled by a conservative local fallback: messages that would have been escalated are quarantined rather than accepted, and the vehicle activates cooperative V2V mode per the distributed misbehavior detection protocol in [13], sharing anomaly observations directly with surrounding vehicles.

### C. *Tier 2 - RSU Edge Agent*

*Role and Scope*: The Tier 2 agent runs on roadside unit infrastructure and is responsible for zone-wide threat assessment and safety-security conflict resolution. It operates within a 10–50 millisecond budget [6], receiving escalated messages and threat signals from multiple vehicles within its coverage zone simultaneously. Unlike Tier 1, it has visibility across vehicles - it can compare what vehicle A reports about vehicle B against what vehicle C independently observes about the same location, providing the cross-vehicle context Tier 1 cannot access.

*Zone-Wide Threat Correlation*: When the Tier 2 agent receives an escalated message from one vehicle, it queries the recent observation logs of other vehicles within the zone for corroborating or contradicting evidence. A Sybil campaign generating phantom vehicles at a specific location will produce BSMs that multiple legitimate vehicles report as physically absent - a pattern invisible to any single vehicle but detectable at the zone level [13]. A GPS spoofing attack affecting one vehicle will produce position reports inconsistent with that vehicle's observed trajectory from other vehicles' perspectives - again, only detectable with cross-vehicle data. The Tier 2 agent aggregates these observations and produces a zone-level threat confidence score for the escalated message.

*Safety-Security Conflict Resolution*: The Tier 2 agent is the only tier with sufficient context to resolve safety-security conflicts. When a message is simultaneously suspicious - exhibiting anomalies suggesting fabrication - and safety-critical - claiming an imminent collision or road hazard - neither dropping nor accepting is straightforwardly correct. The Tier 2 agent resolves this by applying a priority ordering: if the claimed hazard is corroborated by at least one independent vehicle observation within the zone, the message is accepted despite anomalies; if no corroborating observation exists, the message is dropped and a warning is issued to the requesting vehicle. This mechanism converts the safety-security tradeoff from an unresolved design tension into an explicit decision rule with a defined information requirement.

*Output Actions*: The Tier 2 agent produces one of three outputs for each escalated message: Accept with zone-level confirmation, Drop with zone-level rejection, or Fleet-Escalate - forwarding the threat pattern to Tier 3 as a candidate for model update if the threat represents a novel attack pattern not covered by the current detection model.

### D. Tier 3 - Cloud Agent

*Role and Scope*: The Tier 3 agent operates in the cloud on a non-real-time basis, managing fleet-wide model refinement and distributing updated detection weights to all enrolled vehicles. It does not participate in per-message decisions. Its input is the stream of gradient updates and Fleet-Escalate signals from the fleet; its output is updated detection model weights distributed to Tier 1 agents across all vehicles.

*Federated Learning Architecture*: The Tier 3 agent uses federated learning to refine detection models without centralizing raw vehicle data. Each vehicle trains a local gradient update on its recent observations - including novel attack patterns flagged by Tier 1 and Tier 2 - and transmits the gradient, not the underlying data, to the cloud agent. The cloud aggregates gradients from across the fleet and produces an updated global model. Raw sensor data, location history, and message logs never leave the vehicle, preserving both privacy and compliance with automotive data regulations.

The base aggregation algorithm is FedAvg. However, because a coalition of compromised vehicles can inject poisoned gradients designed to degrade detection accuracy, FedAvg alone is insufficient. Informed by evaluations of Byzantine-robust aggregation methods in the autonomous vehicle federated learning literature [12], the Tier 3 agent applies a two-stage defense: Krum selection - which identifies the gradient closest to the geometric median of all received gradients and excludes statistical outliers - followed by Median aggregation across the retained gradients. A single compromised vehicle, or even a small coalition below the Byzantine fault threshold, cannot corrupt the global model through this pipeline.

*Model Distribution*: Updated model weights are distributed to all enrolled vehicles via an OTA mechanism secured by asymmetric cryptography. The model package is signed with a private key held by the framework operator; each vehicle verifies the signature before installing the update, using the same IEEE 1609.2 certificate infrastructure that secures V2X message authentication [10]. A rollback mechanism reverts to the previous model version if a new model produces anomalous classification behavior on the receiving vehicle's local validation set.

### E. Inter-Tier Communication and Trust

The coordination between tiers is facilitated by authenticated communication channels, as summarized in Table 1. Each path is defined by a specific protocol, authentication requirement, and payload structure designed to maintain the chain of trust across the distributed pipeline while mitigating tier-specific threats.

The three tiers communicate through authenticated channels that extend the IEEE 1609.2 PKI infrastructure beyond its standard V2V and V2I scope [9], [10]. Tier 1 to Tier 2 escalation messages carry the original V2X message, the Tier 1 agent's confidence score, the cross-modal consistency check result, and a vehicle pseudonym certificate. Tier 2 to Tier 3 Fleet-Escalate signals carry the zone-level threat pattern, the aggregated observation evidence, and the RSU's identity certificate. Tier 3 to Tier 1 model updates carry the updated weight package and the operator's cryptographic signature.

**Table 1 Inter-tier communication properties, trust mechanisms, and threat mitigation strategies across the three-tier architecture.**

| ***Property/ Comms*** | ***Vehicle → RSU*** | ***RSU → Cloud*** | ***Cloud → Vehicle (OTA)*** |
|---|---|---|---|
| ***Protocol*** | IEEE 1609.2 WSMP DSRC or LTE-V2X | Secure channel (TLS or equivalent) | Secure channel (TLS or equivalent) |
| ***Auth*** | Pseudonym certificate from SCMS | RSU identity certificate | Framework operator-signed model package |
| ***Payload*** | Escalation alerts + sensor snapshots (anonymized) | Aggregated zone threat reports + anonymized updates | Updated ML model weight package + operator's cryptographic signature |
| ***Threat*** | Compromised vehicle | Tampered threat reports | Model poisoning |
| ***Mitigation*** | cert revocation + RSU anomaly scoring | PKI chain-of-trust + BFT aggregation | Signature verification + BFT aggregation |

Each tier verifies the identity of the tier it is communicating with before processing any input. A Tier 2 agent that receives an escalation from a vehicle not enrolled in SCMS discards it. A Tier 1 agent that receives a model update whose signature does not verify against the operator's known public key rejects it. This chain of trust ensures that the inter-tier communication channels do not themselves become an attack surface.

### F. Why This Structure Is Architecturally Forced

A natural question is whether the three-tier structure is necessary, or whether a simpler architecture - a single sophisticated in-vehicle agent, or a two-tier vehicle-cloud design - could achieve the same security properties.

A single in-vehicle agent cannot satisfy the framework's requirements for two reasons. First, it has no access to cross-vehicle observations, making coordinated attack detection impossible regardless of how sophisticated its local classifier is. Second, federated model updates require a coordination layer; without Tier 2 and Tier 3, each vehicle learns independently and the fleet cannot pool observations of novel attack patterns.

A two-tier vehicle-cloud design eliminates Tier 2 and routes all escalations directly to the cloud. This fails on latency grounds: the combined vehicle-to-cloud round trip under realistic 4G/5G conditions can reach 50–200 milliseconds - factoring the 10–50ms server-side latency reported in the literature [6] together with radio access, backhaul, and handover overhead - exceeding the 100-millisecond BSM cycle and making cloud-assisted per-message decisions architecturally impossible. Zone-wide threat correlation requires a compute layer physically close to the vehicles, which is the RSU edge, not the cloud.

The three-tier structure is therefore not a design preference. It is the minimum architecture that simultaneously satisfies the sub-60ms combined pipeline requirement, the cross-vehicle observation requirement, and the fleet-wide learning requirement within the constraints the V2X environment imposes.

## V. Security Analysis

The proposed three-tier multi-agent framework is evaluated against the four attack classes defined in Section II: Sybil attacks, GPS spoofing, replay attacks, and adversarial model poisoning. This analysis describes the structural coverage properties of the architecture, identifying the tier responsible for detection, the response logic, and any residual risks. Table 2 summarizes the structural coverage of the three-tier architecture against each attack class, identifying the tier responsible for primary detection and the role each supporting tier plays in the response pipeline.

### A. Sybil Attack Coverage

A Sybil attacker injecting BSMs from phantom identities creates an anomaly pattern that is architecturally invisible to any per-vehicle system. Because the evidence of a Sybil campaign is distributed across simultaneous observations from multiple receivers, detection cannot be handled solely at Tier 1.

**Table 2 Attack-to-tier coverage mapping showing the primary detection tier (green), supporting tier contributions, and tier applicability for each attack class defined in Section II.**

| | *Tier 1 In-vehicle* | *Tier 2 RSU* | *Tier 3 Cloud* |
|---|---|---|---|
| *Sybil Attack* | Escalate | Primary detection – Fleet alert | Future model update |
| *GPS Spoofing* | Primary detection – Escalate/Drop | Cross-vehicle sensor confirmation | Future model update |
| *Replay Attack* | Primary detection – Drop | Not required | Not required |
| *Model Poisoning* | NA | NA | Primary detection |

The Tier 1 in-vehicle agent performs single-message anomaly scoring against its local behavioral model. If a well-constructed Sybil attack injects individual BSMs that are plausible in isolation, the Tier 1 agent produces an Escalate output due to high uncertainty. This is the framework's conservative bias: when local context cannot resolve a classification, Escalation preserves safety by not dropping a potentially real alert while allowing a higher-tier agent with cross-vehicle context to evaluate it.

At Tier 2, the RSU edge agent aggregates signals from multiple vehicles within a zone. A Sybil campaign becomes statistically detectable at this tier as the spatial distribution of phantom positions is cross-checked against the traffic density and physical sensor returns reported by legitimate vehicles [5], [13]. Tier 2 responses include zone-wide credential flagging and the issuance of a Sybil alert to all local vehicles to suppress BSMs from the flagged pseudonym clusters.

The residual risk is certificate propagation latency. IEEE 1609.2 pseudonym rotation allows an attacker to cycle to fresh credentials before revocation propagates [10]. The framework's behavioral flagging at Tier 2 mitigates this by identifying motion inconsistencies rather than relying on identity alone, though the gap remains a recognized challenge.

### B. GPS Spoofing Coverage

GPS spoofing falsifies the position data in V2X messages, threatening cooperative collision avoidance. Detection requires cross-checking reported coordinates against physical reality - a capability that PKI authentication cannot provide [10].

The Tier 1 agent implements the cross-modal reasoning principle from PhyScout [11]: reported GPS positions are cross-checked against LIDAR, radar, and dead reckoning data. If a BSM reports a neighboring vehicle at a position where no physical signature exists, the inconsistency triggers an Escalate output. The sub-10ms Tier 1 budget limits this to lightweight consistency scoring, but the check is sufficient to justify escalation. For attacks targeting the receiving vehicle's own GPS, the agent detects the divergence between satellite data and internal IMU trajectory.

The residual risk is sensor availability. If a vehicle's sensors are obstructed by weather or urban geometry, the Tier 1 agent loses its physical reference and defaults to Escalate for all position-bearing messages. This preserves safety at the cost of a higher false escalation rate during sensor-degraded operation.

### C. Replay Attack Coverage

Replay attacks retransmit legitimate V2X messages at a different time or location. The primary defense is timestamp and sequence number validation per IEEE 1609.2 [10], which is handled at Tier 1.

The Tier 1 agent checks each message's certificate timestamp and sequence counter. A message that fails timestamp validation is classified as Drop - the only response the framework executes without higher-tier confirmation. This is justified because a message failing timestamp checks cannot be a legitimate safety alert. For replayed DENMs, geographic validity boundaries and event expiry times defined in ETSI EN 302 637-3 provide further disambiguation [19].

The residual risk involves "sub-window" replays where an attacker retransmits a message within the timestamp validity period. While IEEE 1609.2 provides a replay detection service to catch exact duplicates [10], catching sophisticated payload-consistent replays requires additional application-layer plausibility checking as discussed in the standard [10].

### D. Adversarial Model Poisoning Coverage

Model poisoning targets the Tier 3 federated learning pipeline rather than real-time processing. Compromised vehicles submit corrupted model updates to shift the global detection model toward false negatives.

The Tier 3 cloud agent applies Byzantine fault-tolerant aggregation to these updates. Informed by evaluations of Byzantine-robust aggregation methods in the autonomous vehicle federated learning literature [12], the Tier 3 agent applies outlier-resistant aggregation: update vectors are evaluated for consistency with the honest majority, and those that deviate significantly from the consensus distribution are excluded before the global model is computed. This provides robustness against poisoning coalitions up to the fault tolerance threshold. Traceability is maintained through update provenance records, supporting the explainability requirements identified in Section III [8], [14].

The residual risk is the honest-majority assumption. If a coordinated attack compromises a fraction of the fleet exceeding the fault tolerance threshold, the aggregation mechanism cannot guarantee robustness. This scenario requires out-of-band behavioral analysis of fleet-wide detection accuracy trends.

### *E. Cross-Tier Coordination and the Escalation Protocol*

The framework's escalation protocol handles ambiguous signals without requiring the vehicle to determine the attack type. The Escalate response at Tier 1 passes the message, the anomaly score, and the consistency evidence (the metadata derived from the sensor check) to the Tier 2 agent.

This context packaging allows Tier 2 to resolve safety-security conflicts using more information than was available to Tier 1. A BSM that triggered an escalation at Tier 1 due to borderline position inconsistency can be verified at Tier 2 against the independent observations of other vehicles in the zone. If multiple vehicles confirm the reported position, Tier 2 resolves the conflict with an Accept output. If no corroboration exists, Tier 2 issues a Drop resolution and a zone-wide alert. This ensures the safety-security tradeoff is resolved at the tier with sufficient context to make a correct decision.

## VI. Challenges and Open Problems

The three-tier architecture proposed in this paper addresses a set of structural gaps in the V2X security literature. It also inherits open problems that an honest architectural analysis must surface. This section identifies four such challenges and frames them as concrete directions for future work.

### *A. Adversarial Evasion of Cross-Modal Reasoning*

The Tier 1 cross-modal consistency check depends on the assumption that an attacker cannot simultaneously falsify both the V2X message content and the physical sensor observations that would contradict it. For GPS spoofing, PhyScout shows that this assumption holds against attacks that manipulate satellite signals while leaving LIDAR and radar returns intact [11]. It does not hold against a more capable adversary who also manipulates the victim's onboard sensors - through physical tampering, adversarial camera perturbations, or coordinated RF interference against radar.

A sensor-manipulation adversary targeting both V2X messages and the physical observations used to validate them could construct attacks that pass the Tier 1 consistency check by design. Defending against this requires either hardware-rooted sensor attestation - a mechanism that verifies outputs are untampered before entering the detection pipeline - or redundant sensing modalities with independent attack surfaces. The gap analysis of ISO/SAE 21434 identifies sensor integrity as part of the automotive cybersecurity engineering lifecycle [14], but does not specify technical countermeasures at the detection layer.

### *B. Certificate Propagation Latency and the Revocation Gap*

Section V identified certificate propagation latency as a residual risk in Sybil attack coverage. This deserves fuller treatment because it is not an incidental gap - it is a structural tension between two design requirements of the IEEE 1609.2 PKI architecture.

Pseudonym rotation, which is essential for privacy protection, means that a vehicle's credential identity changes frequently [9], [10]. Certificate revocation lists must propagate to every vehicle in range before a revoked credential becomes untrustworthy. In high-mobility V2X environments - highway platooning, intersection approach - a vehicle can enter and leave radio range of a compromised peer before revocation information arrives. The framework's Tier 2 behavioral flagging partially compensates by detecting motion pattern inconsistencies that survive credential rotation, but this is not a complete solution: a sophisticated Sybil attacker whose phantom vehicles exhibit plausible motion trajectories could evade behavioral detection during the propagation window.

Closing this gap likely requires changes to the SCMS architecture itself - shorter certificate lifetimes, faster revocation propagation, or supplementary real-time blocklist distribution through RSU infrastructure - rather than changes to the detection framework layer. This is a standards-level problem the proposed architecture cannot solve unilaterally.

### *C. Regulatory Frameworks for Autonomous Security Response*

The framework proposes autonomous execution of Drop and zone-wide Quarantine responses without human confirmation. This is architecturally necessary given the 100-millisecond BSM cycle constraint, but it creates a regulatory gap with no current resolution.

No existing automotive cybersecurity standard governs what actions an autonomous security system is permitted to take without human authorization. The gap analysis of ISO/SAE 21434 identifies incident response and decision accountability as unresolved in the current standard framework [14]. The SCMS governance model does not address autonomous credential flagging by individual vehicles or RSUs. In the European regulatory environment, ETSI standards define message formats and transmission requirements [18], [19] but are silent on autonomous response authority.

The practical consequence is that deploying the proposed framework at full capability - including autonomous Drop and zone-wide alert issuance - likely requires new regulatory frameworks that do not yet exist. Near-term deployment may need to constrain autonomous action to Tier 1 Escalate-only, with Drop and Quarantine reserved for human-confirmed responses, accepting the latency cost as a regulatory concession. This represents a meaningful reduction in framework effectiveness that future standardization work should address.

### D. Scalability of Federated Aggregation Under Fleet Growth

The Tier 3 Byzantine fault-tolerant aggregation operates under the honest-majority assumption: robustness holds when the fraction of compromised participants stays below the fault tolerance threshold. As fleet size grows, this assumption becomes harder to verify empirically and more consequential when violated.

Two scalability challenges emerge. First, the computational cost of aggregation methods that evaluate pairwise update consistency scales with the number of participating vehicles; at very large fleet sizes, this creates a potential bottleneck in the model update distribution cycle. Second, geographic partitioning of federated aggregation - running separate rounds for regional vehicle populations rather than a global fleet - reduces the observation pool available to detect coordinated poisoning campaigns that span multiple regions. An attacker who distributes a poisoning coalition across geographic partitions could fall below the detection threshold in each partition individually while achieving meaningful global degradation in aggregate.

Hierarchical federated aggregation, where intermediate aggregation at the edge tier precedes full cloud-level aggregation, offers a partial mitigation by introducing a consistency check at the zone level before updates reach the global model. Althunayyan et al. establish the hierarchical FL pattern for vehicular intrusion detection [7]. In the proposed framework, this pattern maps naturally to the existing Tier 2 infrastructure, where RSUs can validate local update distributions before forwarding aggregated gradients to Tier 3. The specific aggregation protocol for this hierarchical case is not specified here and represents a concrete direction for future work.

## VII. Conclusion

The security challenge facing connected vehicle networks is not a shortage of detection techniques. It is a shortage of architectures that deploy those techniques within the constraints V2X actually imposes - a 100-millisecond message cycle, a safety system that cannot tolerate suppressed legitimate alerts, and an attack surface that spans individual vehicles, geographic zones, and fleet-wide learning pipelines simultaneously.

This paper has proposed a three-tier multi-agent framework that treats these constraints as architectural requirements rather than performance targets. The in-vehicle agent classifies every incoming V2X message within a sub-10-millisecond budget, with a conservative bias toward Escalation when local context is insufficient - preserving safety while deferring uncertain decisions to a tier with broader context. The RSU edge agent resolves safety-security conflicts across a geographic zone using cross-vehicle sensor corroboration, operating within a combined pipeline budget that fits inside one BSM cycle. The cloud agent refines detection models through Byzantine fault-tolerant federated learning without centralizing raw vehicle data, distributing updated weights to the fleet on a non-real-time basis that does not touch the per-message decision pipeline.

No individual component of this framework is presented as novel in isolation. Cross-modal sensor fusion for spoofing detection [11], Byzantine-robust federated learning for autonomous vehicle environments [12], and distributed cooperative misbehavior detection [13] each have prior treatments. The contribution is their unification into a coordinated architecture with explicit agent role definitions, standards-derived latency budgets grounded in SAE J2735 and ETSI EN 302 637-2, and safety-security conflict resolution as a first-class design constraint. To the best of our knowledge, no prior framework simultaneously satisfies all three.

Four open problems bound what the framework can currently guarantee. Sensor-manipulation adversaries who target both V2X messages and the physical observations used to validate them can evade Tier 1 cross-modal reasoning - closing this gap requires hardware-rooted sensor attestation that current automotive standards do not yet specify. Certificate propagation latency creates a revocation window that behavioral flagging only partially closes, and resolving it requires changes at the SCMS standards level rather than the detection layer. The autonomous execution of Drop and Quarantine responses operates without regulatory authorization frameworks that do not yet exist, and near-term deployment may require constraining autonomous action to Escalation-only pending future standardization. Finally, Byzantine fault tolerance at fleet scale degrades when poisoning coalitions are distributed across geographic aggregation partitions, and hierarchical aggregation protocols that address this remain unspecified.

These are not objections to the framework. They are the research agenda the framework makes visible. A system that cannot name what it cannot do is not ready for serious evaluation. This one can.

## References

[1] B. Raghavan and B. Schneier, "Agentic AI's OODA Loop Problem," IEEE Security & Privacy, vol. 23, no. 3, 2025. DOI: 10.1109/MSEC.2025.3604105

[2] V. Vinay et al., "The Evolution of Agentic AI in Cybersecurity: From Single LLM Reasoners to Multi-Agent Systems and Autonomous Pipelines," arXiv:2512.06659 [cs.CR], Dec. 2025. [Online]. Available: https://arxiv.org/abs/2512.06659

[3] A. Foundjem, L. N. Tidjon, L. Da Silva, and F. Khomh, "Multi-agent AI framework for threat mitigation and resilience in machine learning systems," *ACM Trans. Softw. Eng. Methodol.*, Jan. 2026, doi: 10.1145/3780095.

[4] R. K. Agrawal, "AI-Driven Secure Vehicular Networks: A Federated Graph Learning Framework for Intrusion Detection and Adaptive Defense in V2X Systems," Zenodo, Apr. 2026. DOI: 10.5281/zenodo.19636618.

[5] H. Morton et al., "Trust-Aware Sybil Attack Detection for Resilient Vehicular Communication," Wiley Internet Technology Letters, Nov. 2024. DOI: 10.1002/itl2.617

[6] IEEE ComSoc Technology Blog, "Agentic AI and the Future of Communications for Autonomous Vehicles (V2X)," Jul. 2025. [Online]. Available: https://techblog.comsoc.org/2025/07/14/agentic-ai-and-the-future-of-communications-for-autonomous-vehicle-v2x/

[7] M. Althunayyan, A. Javed, and O. Rana, "A Robust Multi-Stage Intrusion Detection System for In-Vehicle Network Security using Hierarchical Federated Learning," Vehicular Communications, vol. 49, p. 100837, 2024. DOI: 10.1016/j.vehcom.2024.100837

[8] S. A. Abdel Hakeem et al., "Explainable AI for Federated Learning-Based Intrusion Detection Systems in Connected Vehicles," Electronics, vol. 14, no. 22, p. 4508, 2025. DOI: 10.3390/electronics14224508

[9] A. C. H. Chen, C.-K. Liu, C.-F. Lin, and B.-Y. Lin, "V2X Credential Management System Comparison Based on IEEE 1609.2.1 and ETSI

TS 102 941," in Proc. IEEE VTC 2024-Fall, Washington DC, USA, Oct. 2024. DOI: 10.1109/VTC2024-Fall63682.2024.10774595

[10] IEEE, "IEEE Standard for Wireless Access in Vehicular Environments - Security Services for Applications and Management Messages," IEEE Std 1609.2-2022, 2022. DOI: 10.1109/IEEESTD.2022.9761455

[11] Y. Xu, G. Deng, X. Han, G. Li, H. Qiu, and T. Zhang, "PhyScout: Detecting Sensor Spoofing Attacks via Spatiotemporal Consistency," in Proc. ACM SIGSAC Conference on Computer and Communications Security (CCS), 2024, pp. 1879–1893. DOI: 10.1145/3658644.3670307

[12] J.-H. Chen, M.-R. Chen, G.-Q. Zeng, and J.-S. Weng, "BDFL: A Byzantine-Fault-Tolerance Decentralized Federated Learning Method for Autonomous Vehicle," IEEE Transactions on Vehicular Technology, vol. 70, no. 9, pp. 8639–8652, Sep. 2021. DOI: 10.1109/TVT.2021.3102121

[13] R. Ben Mokhtar et al., "Distributed Misbehavior Detection System for Cooperative Driving Networks," in Proc. IEEE Vehicular Technology Conference (VTC), 2024. DOI: 10.1109/VTC2024-Fall63682.2024.10707235

[14] D. Grimm, A. Lautenbach, M. Almgren et al., "Gap Analysis of ISO/SAE 21434 – Improving the Automotive Cybersecurity Engineering Life Cycle," in Proc. IEEE 26th International Conference on Intelligent Transportation Systems (ITSC), 2023, pp. 1904–1911. DOI: 10.1109/ITSC57777.2023.10422100

[15] G. Asaouer and D. E. Boubiche, "Generative AI-Based Intrusion Detection Systems for Intra-Vehicle Networks," Ad Hoc Networks, vol. 180, p. 104031, Elsevier, Sep. 2025. DOI: 10.1016/j.adhoc.2025.104031

[16] H. Kurunathan, H. I. Ali, G. Javanmardi, M. Eldefrawy, M. G. Gaitán, R. Robles, P. Yomsi, and E. Tovar, "Adaptive Intrusion Mitigation in Software-Defined Vehicles Using Deep Reinforcement Learning," in Proc. 4th Int. Workshop on Real-time and IntelliGent Edge Computing (RAGE '25), Irvine, CA, USA, May 2025, pp. 1–6. DOI: 10.1145/3722567.3727848

[17] SAE International, "J2735: V2X Communications Message Set Dictionary," SAE Standard J2735_202309, Sep. 2023. [Online]. Available: https://www.sae.org/standards/content/j2735_202309/

[18] ETSI, "EN 302 637-2 V1.4.1: Intelligent Transport Systems (ITS); Vehicular Communications; Basic Set of Applications; Part 2: Specification of Cooperative Awareness Basic Service," ETSI EN 302 637-2, Apr. 2019. [Online]. Available: https://www.etsi.org/deliver/etsi_en/302600_302699/30263702/01.04.01_60/en_30263702v010401p.pdf

[19] ETSI, "EN 302 637-3 V1.3.1: Intelligent Transport Systems (ITS); Vehicular Communications; Basic Set of Applications; Part 3: Specifications of Decentralized Environmental Notification Basic Service," ETSI EN 302 637-3, Apr. 2019. [Online]. Available: https://www.etsi.org/deliver/etsi_en/302600_302699/30263703/01.03.01_60/en_30263703v010301p.pdf